\documentclass[twocolumn]{aastex631}
\received{\today}
\revised{\today}
\accepted{\today}
\submitjournal{ApJ}

\graphicspath{{./}{figures/}}
\usepackage{graphicx}	
\usepackage{amsmath}	
\usepackage{amssymb}	
\usepackage{float,hyperref}
\usepackage{color,ulem}
\usepackage{natbib}
\definecolor{darkgreen}{rgb}{0.09, 0.45, 0.27}
\definecolor{darkred}{rgb}{0.90 0.29 0.27}
\definecolor{darkgreen}{rgb}{0.27 0.70 0.50}

\def\refnew#1{(\ref{#1})}

\def\g{\, \rm g}
\def\K{\, \rm K}

\def\cm{\, \rm cm}
\def\m{\, \rm m}

\definecolor{newgreen}{rgb}{0.4,0.9,0.4}

\newcommand{\z}{\color{red}}
\newcommand{\w}{\color{blue}}
\newcommand{\y}{\color{red}}
\newcommand{\x}{\sout}

\defcitealias{CG97}{CG97}
\defcitealias{WL}{WL21}
\defcitealias{MF}{MFK22}
\defcitealias{Pav}{PMA22}

\begin{document}

\shorttitle{Stair-case Disks}
\shortauthors{Kutra et al.}

\title{Irradiated Disks May Settle into Staircases}

\correspondingauthor{Taylor Kutra}
\email{kutra@astro.utoronto.ca}

\author[0000-0002-7219-0064]{Taylor Kutra}

\author[0000-0003-0511-0893]{Yanqin Wu}
\affiliation{
Department of Astronomy \& Astrophysics, University of Toronto}

\author[0000-0003-4450-0528]{Yoram Lithwick}\affiliation{Dept. of Physics and Astronomy, Northwestern University
\& Center for Interdisciplinary Exploration and Research in Astrophysics (CIERA)}

\begin{abstract}
Much of a protoplanetary disk is thermally controlled by irradiation from the central star. Such a disk, long thought to have a smoothly flaring shape, is unstable to the so-called `irradiation instability'. But what's the outcome of such an instability? In particular, is it possible that such a disk settles into a shape that is immune to the instability? We combine Athena++ with a simplified thermal treatment to show that  passively heated disks  settle into a `staircase' shape. Here,  the disk is punctuated by bright rings and dark gaps, with the bright rings intercepting the lion's share of stellar illumination, and the dark gaps hidden in their shadows. The optical surface of such a disk (height at which starlight is absorbed) resembles a staircase. Although our simulations do not have realistic radiative transfer, we use the RADMC3d code to show that this steady state is in good thermal equilibrium. It is possible that realistic disks  reach such a state via ways not captured by our simulations. In contrast to our results here, two previous studies have claimed that irradiated disks stay smooth. We show here that they err on different issues. The staircase state, if confirmed by more sophisticated radiative hydrodynamic simulations, has a range of implications for disk evolution and planet formation.  
\end{abstract} 
\keywords{}

\section{Introduction} \label{sec:intro}
Stellar irradiation is the dominant heat source in most parts of a protoplanetary disk, except perhaps for the inner sub-AU zone \citep[see, e.g.][]{dalessio1998}. In such a `passively heated' disk, small grains in the upper layers absorb stellar photons and re-radiate them as heat. Half of this is lost to space and the other half heats the disk and provides thermal pressure against vertical gravity. 

\citet[][hereafter \citetalias{CG97}]{CG97} proposed a simple power-law solution ('flared disk') for the equilibrium state of these passive disks. This is widely used, both to interpret the spectral energy distribution observed from real systems, and to construct physical models to study planet formation. 

However, the stability of such solutions to thermal perturbations has recently been called into question. Two studies, \citet{Watanabe2008,WL},  argued that passive disks suffer from a linear instability, called the `thermal wave instability' in \citet{Watanabe2008}, and termed more precisely as the `irradiation instability' in \citet[][hereafter \citetalias{WL}]{WL}. It arises because, when an annulus of  a disk is thermally perturbed  and acquires a larger scale-height, its optical surface (the altitude at which stellar irradiation is intercepted) can flare more strongly, allowing it to intercept even more starlight. The perturbation then grows, and propagates inwardly. At the core of the irradiation instability is the amount of stellar flux intercepted by a disk which depends on its vertical structure. 

If this instability indeed operates and can grow into order unity amplitudes, the disk may appear (to the central star) as a `staircase', with steep star-facing edges (`risers') that receive almost all of the stellar flux, interspersed with `treads' that are cast in their shadows. If so, this may explain the formation of gaps and rings commonly observed in resolved protoplanetary disks\citep[e.g.][]{Andrews2018,Huang2018}. It would also affect dust migration and wafting, as well as angular momentum re-distribution in the disk. In short, the irradiation instability may strongly impact the appearances and evolution of   protoplanetary disks.

 At the moment, we do not yet have a full understanding of the instability.  By necessity, the 1-D simulation of \citet{Watanabe2008} and the semi-analytical study by \citetalias{WL} adopted two critical yet problematic assumptions. First, it is assumed that the disk remains vertically isothermal when it is thermally perturbed, while in reality any changes in stellar heating are communicated towards the disk midplane by radiative transfer and can be slow. Second, the disk is assumed to remain in hydrostatic equilibrium at all times. This ignores any possible feedback from hydrodynamics. Such issues are also present in studies by \citet{Siebenmorgen,Ueda2021,Okuzumi2022} As a result, it remains unclear if  the irradiation instability operates under realistic conditions \citep{WL,pav2022}.

In fact, two recent studies by \citet[][hereafter \citetalias{MF}]{MF} and \citet[][hereafter \citetalias{Pav}]{Pav} have cast shadows on the irradiation instability. 
 
\citetalias{MF} simulated the hydrodynamical response of irradiated disks using simplified radiative transfer (the so-called 'moments method'). They reported that a disk with an initial staircase-like profile smooths out into a featureless form after a few thermal times. This lead them to conclude that the irradiation instability is suppressed by the combined effects of vertical thermal diffusion and fluid advection, exactly the two processes ignored by \citet{Watanabe2008} and \citetalias{WL}. Disconcertingly, based on a hydro code with a different radiative treatment, \citetalias{Pav} also reported that the irradiation instability is suppressed in their simulations.

To proceed, it is clear that more realistic treatments of radiative transfer are needed. Unfortunately, sophisticated radiation hydro codes currently under development \citep[see, e.g.][]{Jiang2014,rothkasen} are resource intensive and remain proprietary. So in  this work, we take a different approach to the problem. Instead of investigating the evolution of a disk under the instability, we limit ourselves to the following question: is there a steady state of the disk that is immune to the instability? 

In this work, we will employ the Athena++ hydro code 
\citep{Stone2008,Stone2019} to study irradiated disks, thereby relaxing the assumption of hydrostatic equilibrium. We will calculate the amount of starlight intercepted by a disk annulus self-consistently. However, instead of solving for the vertical transfer of this radiation, we will simply assume that it heats up different heights of the disk at a uniform rate.  This is equivalent to assuming vertical isothermality when the disk is in hydrostatic equilibrium, so we will simply call it the 'vertical isothermal' assumption. 

This assumption is only appropriate 
when the disk is in (vertical) thermal equilibrium. But since we are looking for the steady state, a state that satisfies both dynamical and thermal equilibria, we may be forgiven for adopting such an assumption. In other words, such an approach, while not conventional, is justified by its ends.

We will first describe our physical and numerical setup (\S \ref{sec:physicalsetup}). We then present and analyze in detail the steady states that we obtain (\S \ref{sec:steady}). We ponder  the origin of such a steady state (\S \ref{sec:stall}), before re-examining the claims by \citetalias{MF} and \citetalias{Pav} in \S \ref{sec:critique}. We briefly discuss observable signatures of irradiated disks in \S \ref{sec:observable}, before concluding in \S \ref{sec:conclusion}.

\section{Simulation Setup}\label{sec:physicalsetup}

\subsection{Irradiation and Thermal Physics}
\label{subsec:heat}

Here, we discuss the key thermal physics inputs for our Athena++ simulations.

First, some terminology.  Small dust grains at the optical surface intercept stellar photons -- we call these `visual/optical light' -- and re-radiate the heat isotropically. We call the latter radiation `infrared light'. Roughly half of it is lost to the space above and the other half travels downwards towards the disk midplane. The latter half is then absorbed by grains (mostly larger ones) and re-processed into an even colder blackbody radiation -- we call this `mm light', though the wavelength is typically shorter than a millimetre. Unless the opacity law is very steep (see Appendix \ref{sec:opacity}), much of the disk below the optical surface is largely isothermal when in thermal equilibrium.  

We compute the optical light interception as follows.
Let $\tau_r$ be the radial optical depth to stellar photons,
\begin{equation}
\tau_r (r,\theta,\phi) \equiv  \int_{0}^r \rho(r',\theta,\phi) \, \kappa_V dr' \, ,
\label{eq:tau}
\end{equation}
and $\kappa_V$ is the Planck opacity for starlights ($V$ for visual), $\rho$ is the gas density. 

In the analytical work of \citetalias{WL},  stellar heating is determined based on a quantity called the `optical surface', $H(r)$, disk height at which $\tau_r = 1$.\footnote{By definition, $H(r)/r$ cannot decrease with radius. The region of the disk that satisfies $d(H/r)/dr=0$ is in shadow.}
Numerically, it is instead much easier to compute the heating as follows.
The stellar flux at every point is
\begin{equation}
    F_{\rm irr} (r,\theta,\phi) = \left(\frac{R_*}{r}\right)^2 \, \sigma_{\rm SB} T_*^4\, e^{-\tau_r} \, ,
    \label{eq:firr}
\end{equation}
where $\sigma_{\rm SB}$ is the Stefan-Boltzmann constant, and $R_*$ and $T_*$ the stellar radius and surface temperature, respectively. One can then determine the amount of light intercepted by taking the divergence of the flux. In addition to numerical expediency, such a procedure naturally incorporates the case of a `translucent optical surface' whereby starlight is deposited over a finite horizontal distance (this distance is called `smearing length' in \citetalias{WL}). 

We now turn to thermal processing. Any change in heating is communicated to the disk below by radiative transfer, and this proceeds diffusively in the optically thick region. So a perturbed disk is not expected to remain vertically isothermal. However, as we are interested in the final steady state, we will continue making the simplifying assumption that different vertical layers of the disk 
are heated up in unison.
Let us define a quantity called the `forcing temperature', based on the height-integrated heating at a given radius,
\begin{equation}\label{eq:tforce}
    \sigma_{\rm SB} T_{\rm  force }^4(r) = \frac{1}{2} \int_0^{z_{\rm max}} (-\nabla\cdot F_{\rm irr}) \,dz \, ,
\end{equation}
where the height $z \equiv r \sin(\theta)$.
We can accurately evaluate this integral even with a small number of vertical grids.

We stipulate that, under radiative forcing, the local temperature, $T(r,\theta)$, relaxes towards the height-independent $T_{\rm force}(r)$ as
\begin{equation}
4 \sigma_{\rm SB} T^3 \frac{\partial {T}}{\partial t }{\bigg \vert}_{\rm rad}= {1\over{\tau_{\rm th}}}\,  \left[ \sigma_{\rm SB} T_{\rm force}^4(r)- \sigma_{\rm SB} T^4\right]\, ,
\label{eq:dtdt}
\end{equation}
where $\tau_{\rm th}$ is the thermal relaxation time (see below),  and $\sigma_{\rm SB} T^4$ is the rate of black-body cooling.
We call this the `vertical isothermal' assumption. The actual temperature may depend on height slightly, due to the presence of compressional heating in the energy equation (eq. \ref{eq:energy}). In practice, we further set the cooling term (the last term on the right hand side) to be $\sigma_{\rm SB} T_{\rm mid}^4$, where $T_{\rm mid}$ is the midplane temperature. This makes little difference to the results. 

\subsection{Numerical Setup}\label{sec:setup}

We use Athena++ \citep{Stone2008,Stone2019}, a grid-based, high-order Godunov code, to conduct 2D (axi-symmetric) hydrodynamic simulations.
Athena++ integrates the following equations of gas dynamics, 
\begin{equation}\label{eq:mass}
    \frac{\partial\rho}{\partial t} + \nabla \cdot [\rho\mathbf{v}] = 0 \,,
\end{equation}
\begin{equation}\label{eq:mom}
    \frac{\partial(\rho\mathbf{v})}{\partial t} + \nabla \cdot [\rho\mathbf{v}\mathbf{v} +P] = -\rho\nabla\Phi \, ,
\end{equation}
\begin{equation}\label{eq:energy} 
    \frac{\partial E}{\partial t} + \nabla \cdot (E+P)\mathbf{v} = - \rho(\mathbf{v}\cdot\nabla\Phi) + \frac{\rho}{\Gamma-1}\frac{k_b}{\mu m_H} \frac{\partial T}{\partial t }{\bigg \vert}_{\rm rad}\, .
\end{equation}
Here  $P$ is gas pressure, $v$ the 3-D velocity, and $\Gamma$ the adiabatic index (we take $\Gamma = 7/5$). The gravitational potential of the central star with mass $M_*$ is given by $\Phi = -GM_*/r$. 
The energy density, $E$, is the sum of internal ($P/(\Gamma-1)$) and kinetic ($\rho v^2/2$) energies. The last term on the right hand side of eq. \refnew{eq:energy} accounts for radiative heating and cooling, and is evaluated using eq. \refnew{eq:dtdt}.
The integration timestep is determined by a minimum Courant number of $0.3$.

\subsubsection{Disk Model} \label{sec:disk}

We choose the following set of parameters for our fiducial disk.

The central star is Sun-like, with solar mass, solar radius and solar  temperature. The initial surface density of the gas disk runs as
\begin{equation}
    \Sigma(r) = \Sigma_0 \Big(\frac{r}{r_0}\Big)^{-1} \, ,  
\end{equation}
with a two-sided surface density $\Sigma_0=1700\, {\rm g/cm}^2$ at $r_0=1 \rm AU$. The initial radial temperature profile is taken from \citetalias{CG97}: 
\begin{equation} \label{eq:CG97} 
    T(r) = T_{\rm CG97} = T_0 \Big(\frac{r}{r_0}\Big)^{-3/7} \, , 
\end{equation}
with  $T_0=150 \K$. The isothermal sound speed is
\begin{equation}
    c_s(r) = \sqrt{\frac{k_b}{\mu m_H}T(r)} \, ,
\end{equation}
    while the disk scale height $h(r) = c_s(r)/\Omega(r)$, with $\Omega(r) = \sqrt{GM/r^3}$ being the Keplerian orbital frequency.
We initialize the gas density assuming vertical hydrostatic equilibrium for a locally isothermal disk. 
The initial meridional velocities ($v_r$, $v_\theta$) are set to zero, and the azimuthal velocity is determined by force balance between gas pressure, gravity and centrifugal force.

For computing the optical depth to the star (eq. \ref{eq:tau}), we adopt a gas opacity of  $\kappa_V=1 {\rm\, cm}^2/{\rm g}$. This is what one expects if about $10^{-4}$ of the gas mass (and therefore $\sim 10^{-2}$ of the dust mass) is in small, micron-sized grains. We ignore scattering of photons.

\begin{figure}
    \centering
    \includegraphics[width=0.45\textwidth]{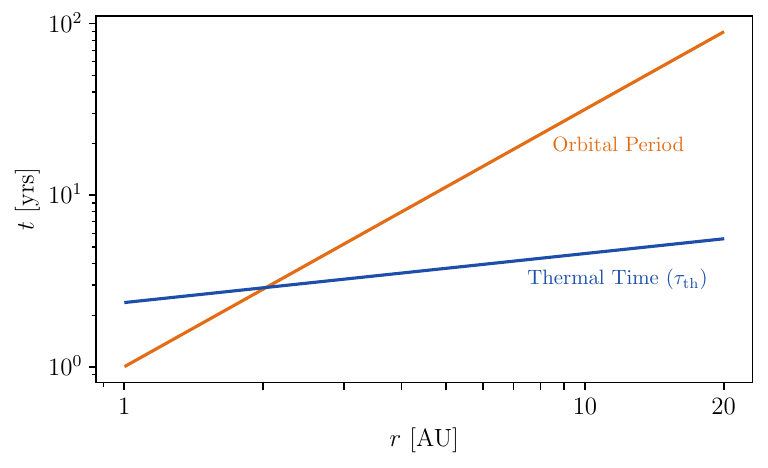}
    \caption{Two timescales, the thermal time (eq. \ref{eq:tauthermal}
    ), and the dynamical time, for our fiducial disk, at its initial state. The thermal time is relatively constant across all radii. This plot indicates that, if a steady state exists, we should be able to reach it within a few hundred years.  }
    \label{fig:tau}
\end{figure}

We emulate \citetalias{WL} and adopt a relaxation time 
\begin{equation}
    \tau_{\rm th} = \frac{3}{32}\frac{c_v \Sigma(r)
    }
    {\sigma_{\rm SB} T^3} \, ,
    \label{eq:tauthermal}
\end{equation}
where $\Sigma(r)$ is the disk surface density, and the specific heat $c_v=\frac{5}{2}\frac{ k_b}{\mu m_H}$  with a mean molecular weight of $\mu=2.3$.  The resultant relaxation time is plotted in Fig. \ref{fig:tau} and it is roughly flat with radius. This timescale controls the speed towards equilibrium. Our choice (eq. \ref{eq:tauthermal}) is shorter than that in \citetalias{WL} and is adopted for computational expediency. We confirm below that its value does not impact the final steady state. 

\subsubsection{Domain and Boundary Conditions}\label{sec:bc} 

Our 2-D hydro simulations are conducted in spherical polar coordinates ($r, \theta$). In the meantime, the relevant thermal physics is best described in a cylindrical grid.
As our disks are thin ($z= r \sin\theta \ll r$), we have simply equated $r$ with the cylindrical radius throughout the paper. For instance, the vertical integration of radiative forcing (eq. \ref{eq:tforce}) is numerically performed over grids of the same spherical radius; similarly, the same heating term (eq. \ref{eq:dtdt}) is applied to cells of the same spherical radius.

Our integration domain is $r\in[1,20] \rm AU$,  $\theta \in [\pi/2, \pi/2-0.3]$ (above the disk mid-plane, where $\pi/2$ is midplane). We  adopt grid numbers $N_R\times N_\theta = 256\times128$. 

The radial grids are logarithmic, with the  spacing expanding outward as $(r_{\rm i+1}-r_{\rm i})/(r_{\rm i}-r_{\rm i-1}) = 1.01$. The boundary conditions for the inner and outer radial boundaries  are that all variables (gas bulk properties and velocities) are held to their initial values. 

For our fiducial run, we follow \citetalias{MF} in assuming the presence of an un-modelled disk inward of our radial domain. This modifies the radial optical depth as
\begin{equation}
\tau_r (r,\theta) \equiv  \tau_0 (\theta) + \int_{r_{\rm in}}^r \rho(r',\theta) \kappa_V dr' \, ,
\end{equation}
where we take $\tau_0 (\theta)  = \kappa_V \times \rho (r_0, \theta) \times (r_0-10R_*)$ \citepalias[see][]{MF}. Such a procedure, together with our radial boundary condition that $\rho(r_0, \theta)$ is constant, erects a static opaque 'screen' at the inner boundary. For our fiducial parameters, this screen blocks starlight up to a height $z/r \approx 0.11$. It prevents overheating of the simulated disk  due to direct exposure. In real disks, the inner material will also adjust to the stellar heating. We investigate one such case in \S \ref{sec:params}. 

We now turn to the $\theta$ direction. Our grids only cover a part of the meridional plane above the equator. We employ a reflecting boundary condition for fluid velocities at $\theta = \pi/2$. This ensues symmetry about the midplane. Our upper boundary at $\theta = \pi/2 - 0.3$ is chosen such that we cover at least 6 pressure scale heights across all radii (scale height is the largest at the outer boundary, $h/r \sim 0.05$). This guarantees that the optical surface ($H$) always lies within our computation domain. 

However, such a cautious choice for the upper boundary introduces a unique computational challenge. 
As gas density drops super-exponentially with height, we inevitably encounter the gas density floor (set at $10^{-18} \g/\cm^3$), the minimum gas density introduced in the code to deal with vacuum. Above this height, gas is assigned a constant density and so cannot stay in hydrostatic equilibrium.  Waves are continuously excited. Moreover, if densities in the upper ghost-cells are held to their initial values, steep pressure gradients may develop when the gas below evolves. This drives shocks. To minimize these two effects,  we choose to set the gas density in the upper boundary to be a linear extrapolation of $\log \rho$ below it. We hold velocities to their initial values. Some waves still persist and we examine their impacts on the simulation in Appendix \ref{app:bc}.

\subsubsection{Horizontal Radiative Transfer}\label{sec:smoothing}

\citetalias{WL} pointed out that there is also heat transfer in the radial (horizontal) direction. There are two separate physical effects. 

The first effect is due to finite disk thickness. Consider midplane gas at radius $r$.  It experiences heating from grains at altitude $H$ directly above it (eq. \ref{eq:tforce}), but it is also sensitive to those within a radial distance $\delta r \leq H$. In other words, the isotropic re-radiation from the small grains at $H$ are felt over a finite radial distance $\delta r \leq H$. Accounting for this effect suppresses thermal  perturbations with very short wavelengths. This is a more significant issue in the outer disk where $H/r$ is larger.

To accommodate this effect, we introduce a horizontal smoothing to the heating function,
\begin{equation} \label{eq:smooth}
    \sigma_{\rm SB} T^4_{\rm force}(r_j)\Big|_{\rm smooth} = \frac{\sum_{i=1}^{N_r} K(r_j, r_i) \sigma_{\rm SB} T_{\rm force}^4(r_i)}{\sum_{i=1}^{N_r} K(r_j, r_i)}\, ,
\end{equation}
with a Gaussian kernel 
\begin{equation} \label{eq:kernel}
    K(r_j,r_i) = \exp\left\{{-{1\over 2} \left[\frac{r_j-r_i}{\beta H(r_j)}\right]^2 }
    \right\}\, .
\end{equation}
Here, $\beta$ is a free parameter and it is expected to be of order unity based on the above physical discussion. We take $\beta=0.4$ for our fiducial case (so that the FWHM is $\sim H(r_j)$) and experiment with different values in \S \ref{sec:params}. 

Another related effect  takes place when the perturbation amplitudes are large. Consider an annulus that is not exposed to direct starlight (i.e., in shadow). It cools and contracts vertically. But it will not reach  zero temperature. Instead, because its   neighbouring rings (which can be at a distance $\delta r \gg H$) are hot and puffed up, IR light from these rings can reach the shadowed zone and keep it illuminated \citep[e.g., see Fig. 1 of][]{Okuzumi2022}. This effect is called ``back-warming'' \citep[e.g.,][]{Dullemond2010}. To emulate this qualitatively, we impose a minimum temperature in the disk that is $45\%$ of $T_{\rm CG97}$ (eq. \ref{eq:CG97}). This value is motivated by results of our RADMC simulations (\S \ref{subsec:radmc}).

Our approaches to solve the two horizontal transfers are crude, and one is referred to \citet{Okuzumi2022} for some more sophisticated treatments. Fortunately,
as we show in \S \ref{sec:params}, the irradiation instability is not suppressed by horizontal transfer, and  our steady states are not qualitatively affected by the details of these treatments. 

\begin{figure}
    \centering
\includegraphics[width=0.47\textwidth]{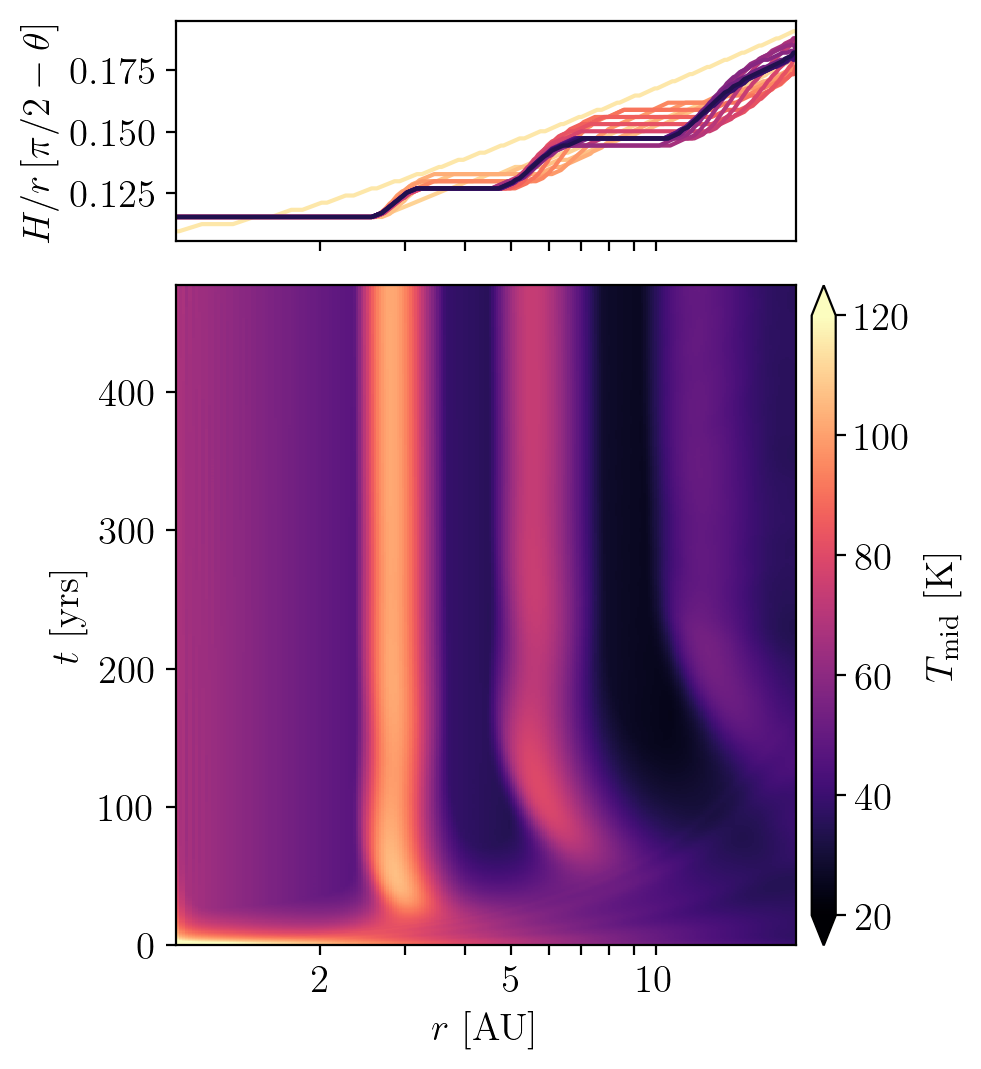}
    \caption{How our fiducial disk reaches its staircase steady state. The top panel shows the height of the optical surface as a function of radius ($H/r$;  snapshots taken every 16 yrs, with initial being the lightest in color; values are slightly smoothed from machine output for aesthetics). At the final equilibrium, the optical surface resembles a series of staircase steps. The bottom panel shows the midplane temperature as a function of time. Starting from an initial power-law state, thermal perturbations develop, move inward and stall, segmenting the disk into hot and cold rings. The inner disk reaches equilibrium earlier. There is a shielding screen that extends up to $z/r \approx 0.11$ at the inner boundary. }
    \label{fig:evol}
\end{figure}

\begin{figure*}[t]
    \centering
\includegraphics[width=\textwidth]{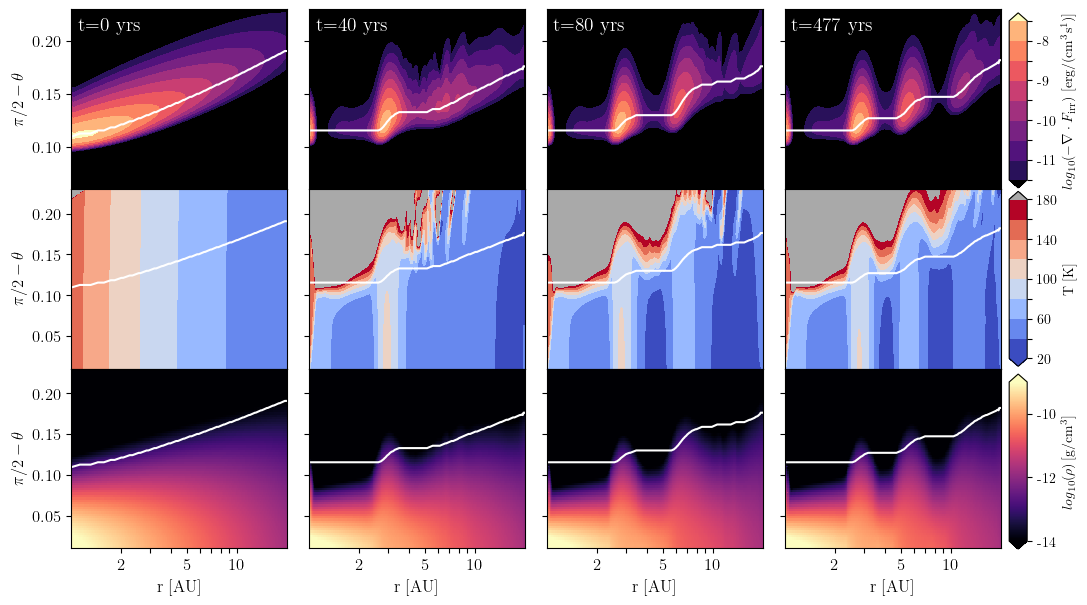}
    \caption{
    Similar to Fig. \ref{fig:evol} but with more details. Time runs from left to right. Each panel shows the 2-D structure of a physical quantity: the left panels are the divergence of the stellar flux (in logarithmic scale), or where irradiation is absorbed by disk; the middle ones are the gas temperature, and the right panels are the gas density (in logarithmic scale). The white curve in each panel indicates the optical surface ($H$). The disk, starting from a power-law temperature profile, is transformed into a 'staircase' after a few thermal times, with most of the stellar heating concentrated near the stair-risers, while regions in-between are cast in shadow (``stair-treads"). The evolution is inside-out. By design, most of the disk is vertically isothermal, except for the low density gas at high altitudes (mostly above the optical surface) that is also affected by compressional heating of boundary-driven waves. This is not consequential, because temperature in this rarefied gas does not control the amount of light the disk intercepts.  An animated version of this simulation can be found in the online article. The 9 second animation starts at $t=0$ years and runs through $t=525$ years and shows the evolution of thermal peaks. The static figure consists of 3 snapshots of this evolution. We have included the temperature, density, $H$, and stellar flux ($F_{\rm irr}$ or eq. \ref{eq:firr}) in supplementary data.  
     }
    \label{fig:fiducial} 
\end{figure*}

\begin{figure*}
    \centering
    \includegraphics[width=0.98\textwidth]{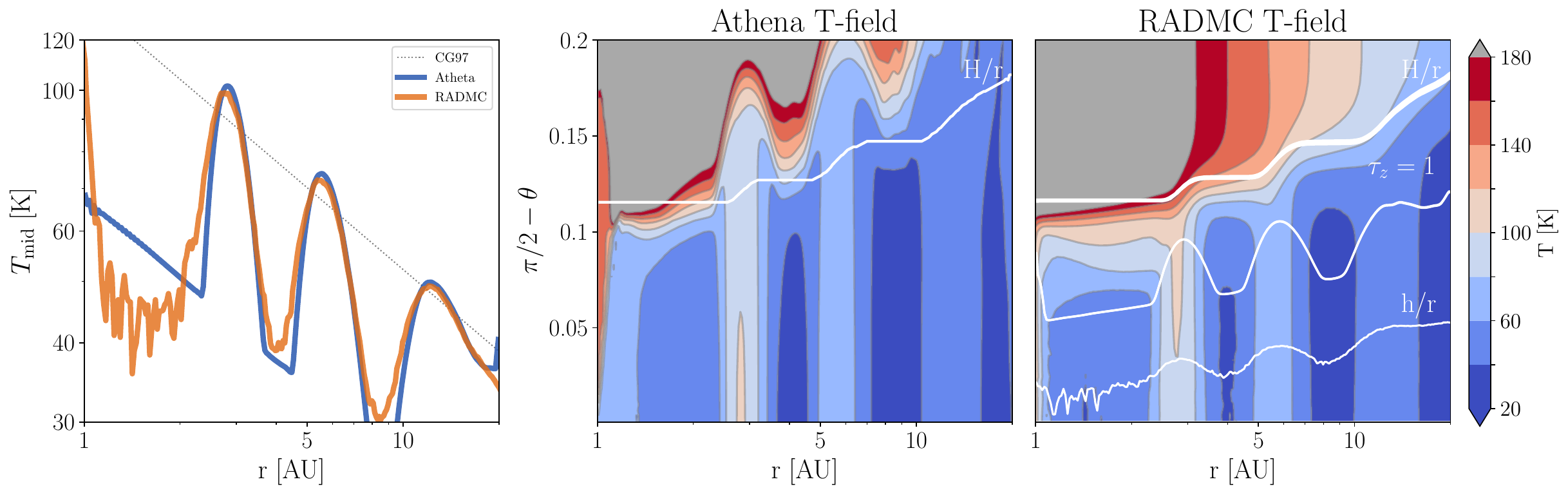}
    \caption{Examining the steady state (labelled as `Athena') using RADMC3d. The left panel compares the midplane temperature outputs from Athena and RADMC3d, with the dotted light line ('CG97') indicating the initial temperature profile (eq. \ref{eq:CG97}).The right two panels show the 2-D temperature fields from Athena (middle) and from RADMC (right). For regions below the optical surface, our Athena steady state share similar temperatures as that obtained from RADMC3d, confirming that our steady state is in good thermal equilibrium. The various labelled curves represent: $H$, the optical surface; $\tau_z = 1$, the disk vertical photosphere; $h$, vertical scale height.  
    }
    \label{fig:compare-radmc}
\end{figure*}

\subsection{Evolution Towards a Steady State}\label{sec:effects} 

The time evolution of our fiducial disk is shown in Fig. \ref{fig:evol}. Starting from a smooth power-law temperature profile, the disk quickly develops temperature kinks that grow in amplitudes as they propagate inward,  at a timescale of the local thermal time. These behaviour are as predicted by the linear theory of \citetalias{WL}. 

In a bit more detail, we see that the region just beyond 1AU cools with time from an early start. This is because it finds itself partially blocked from starlight by the (numerically erected) inner screen that extends to a height of $z/r  \approx 0.11$. The IR light that the midplane receives now is not sufficient to keep it poking above the screen. It has no choice but to cool and contract vertically.\footnote{In this shadowed region, $H/r$ remains constant ($\approx 0.11$) as is set by the inner screen.} Interestingly, its loss is someone else's gain. The region near 3AU now has a relatively unhindered view to the central star,  compared to the case in a power-law disk. This region then heats up and extends vertically. A hot ring is formed. Such a ring is analogous to the so-called `puffed-up inner rims' found in the inner edges of protoplanetary disks  \citep{Natta2001,Dullemond2001}, except here it is in the middle of a disk.

According to the analysis in \citetalias{WL}, such a hot ring should continue to propagate inward. To our surprise, it stalls after travelling only a small distance. This is a key result from our study. We will discuss more below (\S \ref{sec:stall}).

With this ring now in place and taking over the role of the numerical inner screen, the story repeats itself. One after another, hot rings form successively outward. Such an inside-out pattern formation is not surprising, since the inner part of the disk controls the irradiation condition for the outer part.\footnote{The thermal relaxation time in our fiducial case increases outward. However, we confirm later that even for a case where it decreases outward, the behaviour is similar.}
After $\sim 100$ yrs, a total of  three hot rings are established, fringed by cold belts that lie in their shadows.
These features are broad and widely spaced, with widths and spacings all of order the local radius. 

Further integration to beyond $500$ yrs returns no  appreciable changes. In such a steady state, the total amount of stellar light intercepted by the disk is comparable to that of a conventional power-law disk, but the distribution of this light is highly inequitable. The hot rings, with their steep star-facing edges, capture almost all of the starlight (top panels, Fig. \ref{fig:fiducial}). The final disk shape, in terms of the optical surface, resembles a staircase. 

\section{The Steady state}
\label{sec:steady}

This new steady state is our most important result. We now analyze this state in detail, in order to establish its credibility. This is important, especially because, to arrive at it, we have adopted a questionable assumption ("vertical isothermal"). In particular, we ask whether the staircase disk is truly in thermal and dynamical equilibria. After this, we  return to discuss why such a steady state may be immune to further irradiation instability.

\subsection{Inspection using RADMC}
\label{subsec:radmc}

We study the question of thermal equilibrium using the code
RADMC-3d \citep{radmc}. This is a Monte-Carlo photon code that calculates the equilibrium temperature field, for a given density field. We will compare RADMC results against the  temperature field we reach via Athena++. A good agreement indicates that our steady state is in thermal equilibrium.

To conduct RADMC experiments, we have to adopt an opacity law. We choose the following grey opacity, 
\begin{equation}
\kappa_\nu = 1\,  {\rm cm^2}/{\rm g-gas}\, .
\label{eq:kapparadmc}
\end{equation}
Such a form crudely represents that arising from a mixture of 
micron-sized and mm-sized grains \citep[see, e.g.][]{woitke2016}. We choose this opacity law for two reasons. First, it is the same as our choice of visual opacity ($\kappa_V$) in the simulations. Second, a  grey opacity disk should be vertically isothermal at equilibrium (see Appendix \ref{sec:opacity}), therefore compatible with our key assumption. 

As in our hydro simulations, we adopt solar parameters for the central star and suppress dust scattering. The gas density field is taken from the last snapshot ($t = 477$ yrs) in our  fiducial run. We extend this density field both inward and outward slightly.  The inner extension produces the inner screen  described in \S  \ref{sec:bc}; and the outer extension helps to avoid excessive cooling at the outer edge due to a truncated disk. We use $5\times 10^8$ photon packets.
 
The RADMC results are presented in Fig. \ref{fig:compare-radmc}, alongside our Athena++ results. The midplane temperatures of the two simulations agree well, and the vertical temperature distributions are also largely similar, especially for regions below the optical surface. These agreements validate a number of procedures in the hydro simulations, including the deposition of stellar flux, the vertical sharing of this flux, the horizontal smoothing, and the minimum temperature floor that accounts for `back-warming'. 

There exist, however, two discrepancies. 

The first one concerns  temperatures at high altitudes. They are very discrepant between the two codes. This results from two different effects. First, in hydro simulations, the high altitude region is continuously plagued by waves that emanate from the upper boundary (\S \ref{sec:bc}). But since this is well above the optical surface, we believe it is irrelevant to the dynamics. Second, in RADMC, gas at or above the optical surface can see the star directly and is thus  heated to a higher temperature than that in the midplane. This is not captured by our 'vertical isothermal' assumption. Again, we believe this does not strongly impact the dynamics.

The second discrepancy lies in the midplane temperature (the left panel of Fig. \ref{fig:compare-radmc}) and contains two separate sets of features. The first set is the region inside 2AU, where the RADMC result is hotter at  1AU and colder at 1-2AU, compared to the Athena++ ones. The former arises  because the  1AU zone in RADMC is in radiation contact with hotter material inward of it  that is not properly modeled in Athena++. The latter is because our floor temperature treatment ($45\%$ of $T_{\rm CG97}$) over-estimates the effect of back-warming in the 1-2AU neighbourhood. Neither of these is important for the overall dynamics, because this region is in shadow and does not affect the stellar irradiation the outer disk receives. 

The second set of features concerns the annuli immediately inward of each of the three hot rings. While our adoption of a floor temperature seems to work largely, it is not perfect. The biggest offense shows up in these annuli,  where RADMC predicts a higher temperature than Athena++. This discrepancy is worrisome because the physical state of these regions may impact the propagation or stalling of thermal waves. This should be investigated in the future.

Overall, we conclude that our steady state is  in good thermal equilibrium. 

\subsection{Dynamical Equilibrium and Meridional Circulation}\label{sec:meridional}

\begin{figure}
    \centering    \includegraphics[width=0.5\textwidth]{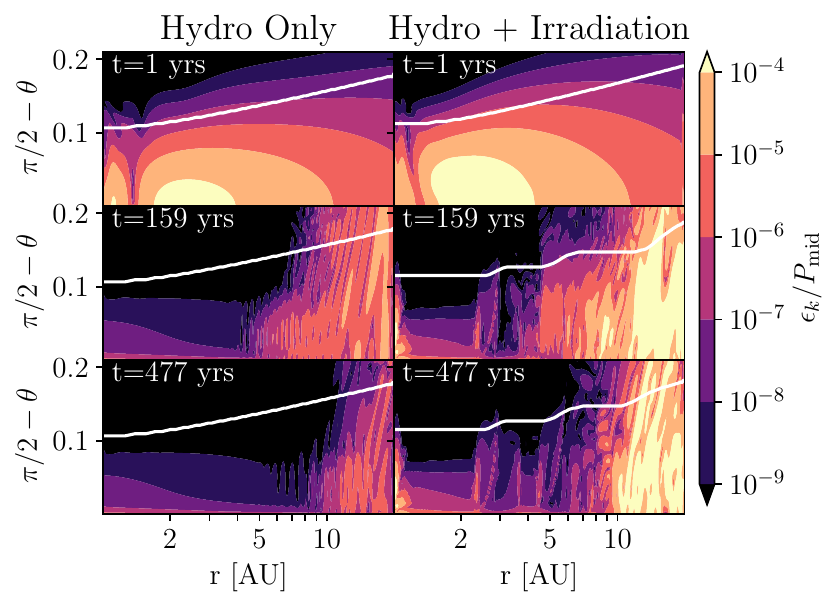}
     \caption{The path towards dynamical equilibrium. Here, we display the ratio of kinetic energy density (eq. \ref{eq:energy}) and the local midplane pressure,  in the 2-D plane, and for both a pure hydro disk (no stellar heating, left) and our fiducial disk (right). Over time (from top to bottom), both disks settle down to a dynamical equilibrium, and the kinetic energy density largely vanishes. The irradiated disk maintains a small meridional circulation near the hot rings. See Fig. \ref{fig:streamlines} for more details. 
    }
    \label{fig:bc}
\end{figure}

\begin{figure}
    \centering    \includegraphics[width=0.5\textwidth]{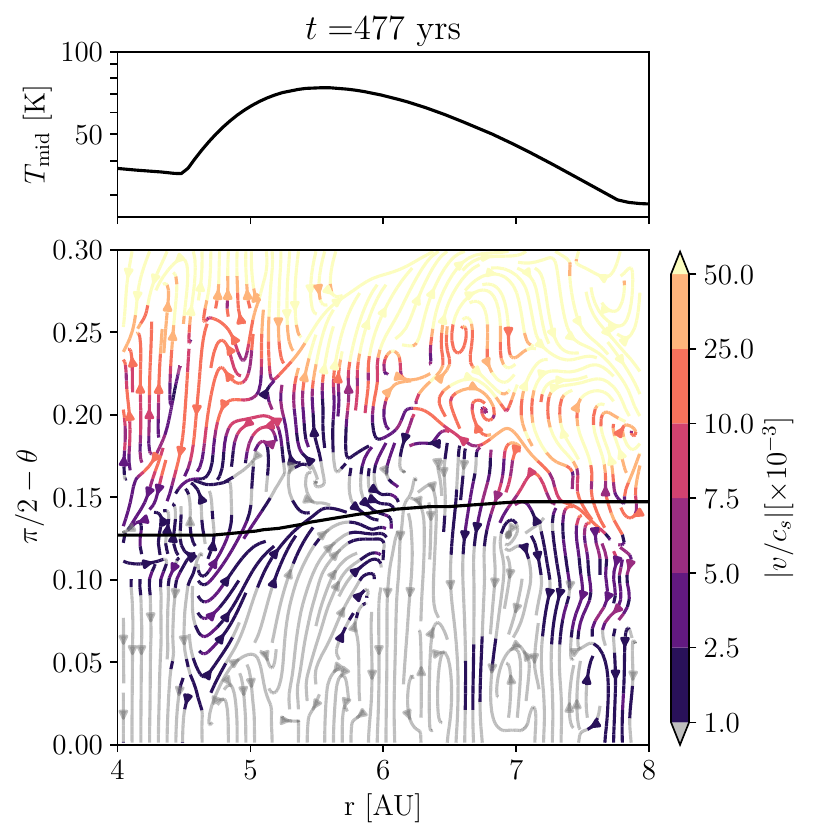}
    \caption{Details around a hot ring for our fiducial disk at steady state. The top panel displays the midplane temperature, and the lower one the  meridional streamlines (scaled by local sound speed), with a  black curve indicating the optical surface ($H$).  The velocities are in general small ($\leq 10^{-3} c_s$), except near the upper boundary. Waves are continuously excited there by the imperfect boundary condition. We have included the meridional and azimuthal velocity profiles as supplementary data.}   \label{fig:streamlines}
\end{figure}

To quantify dynamical equilibrium, we  define a kinetic energy  density for the meridional motion,
\begin{equation}
    \epsilon_k = \frac{1}{2}\rho(v_r^2+v_\theta^2) \, ,
\end{equation}
where $v_r$ and $v_\theta$ are the radial and vertical velocities, respectively. Figure \ref{fig:bc} shows $\epsilon_k$, in ratio to the local midplane pressure. This ratio informs the degree of departure from hydrostatic equilibrium. For comparison, we also integrate a purely hydro disk, one without radiative forcing (the last term in eq. \ref{eq:energy}).

Comparing the pure hydro disk and our fiducial disk (Fig. \ref{fig:bc}), we observe a common trend and some differences. We see that as the two disks adjusts hydrodynamically, their kinetic energy densities decay over time by a few orders of magnitude. This process is faster in the inner region where the dynamical time is shorter. By the end of our simulations (477 yrs), both disks have largely reached dynamical equilibrium, with the ratio of $\epsilon_k/P_{\rm mid}$ falling to $10^{-4}$ or lower, for most of the disk. 

Compared to the pure hydro disk, the irradiated disk takes longer to settle down. This is because the latter is susceptible to unstable thermal waves. Moreover, even after the  disk temperature has reached a steady state, there still remain a small, yet discernible meridional circulation near the hot rings. 

In Fig. \ref{fig:streamlines}, we zoom in to study one such circulation. The streamline plot shows that gas within a couple scale heights of the midplane is largely static, but 
there remains fast, fluctuating, fluid motion at high altitudes. The latter may result, as we explain previously, from the imperfect numerical treatment near the upper boundary. Fortunately, as the pure hydro run in  Fig. \ref{fig:bc} testifies, this issue is not significant for the overall dynamics. We further elaborate on this point in Appendix \ref{app:bc}.

A more interesting feature in the streamline plot is the circulation at moderate latitudes. We observe a weak circulation, with velocity magnitudes of order $10^{-3} c_s$. At this speed, the fluid flow is not competitive against radiative heating/cooling and can at best advect only a small amount of the stellar heating. But it may be important for stirring up the dust grains.
 
In summary, in our 2D simulations, after a few hundred inner orbits, the disk has largely settled down to a dynamical equilibrium, modulated by weak circulations near the hot rings. 

\subsection{The Steady State and Parameter Choices} \label{sec:params}

\begin{figure*}
    \centering
    \includegraphics[width=\textwidth]{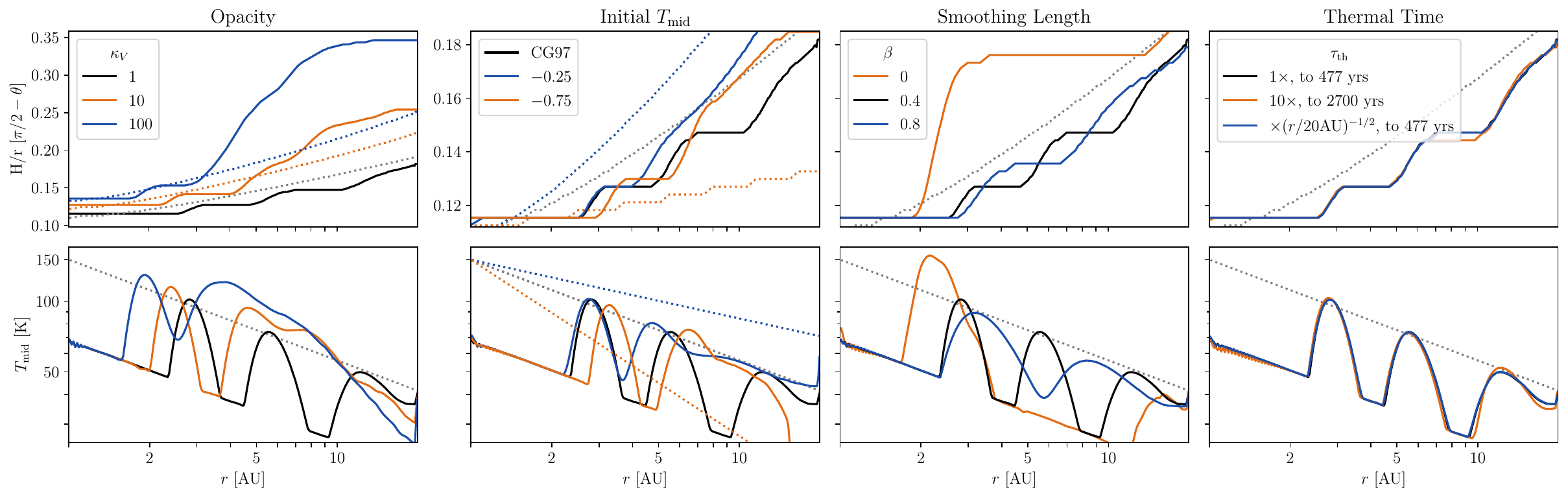}
    \caption{How the final steady state depends on various model parameters. The top panels show the heights of the optical surfaces, and the lower ones the mid-plane temperatures, with the initial states as dotted lines.   By 477 yrs (duration of all runs, except for one case in the right panel), a steady state is reached in every simulation.  Compared to the fiducial run (shown in all panels as a black curve), changes to the opacity  $\kappa_V$ (leftmost column), the initial temperature profile (second to the left column), the smoothing length $\beta$ (third column), and the thermal relaxation time (rightmost column), modify the widths and the spacing of the hot rings at steady state, but do not affect its general character.
    }
    \label{fig:comparison}
\end{figure*}

To gain more insights on the steady state, we study how it depends on our choice of parameters. In each of the following simulations, a staircase like steady state is reached (Fig. \ref{fig:comparison}). The locations and widths of the hot rings depend on the parameters, but not always in ways that are easily interpretable, possibly because these are results of nonlinear evolution.

The first parameter to consider is the optical opacity, $\kappa_V$.  Holding all other parameters fixed, we increase the opacity from $1$ (the fiducial case), to $10$ and to $100 \, {\rm cm}^2/{\rm g}$, corresponding to increasing dustiness. It appears that more opaque disks tend to harbour wider hot rings. This may be due to the higher optical surface (and therefore larger horizontal smoothing, see eq. \ref{eq:kernel}) in more opaque disks. 

We then explore varying the initial condition, in particular, the power-law index in eq. \refnew{eq:CG97} for the initial temperature profile. This appears to alter the final state. This suggests that there is an infinite set of distinct final states the disk can settle down to -- a hot ring can form at any location, as long as it is able to intercept enough sunshine to maintain its vertical height. The initial condition affects the evolutionary path and therefore the nonlinear outcome.

We also  study the impact of horizontal smoothing, by varying the value of $\beta$ in  eq. \refnew{eq:kernel}. 
While cases with larger smoothing lengths tend to harbor broader (and fewer) hot rings, the case with $\beta = 0$ (no smoothing) reaches an unexpected final state. Almost the entire stellar flux is now intercepted by one very hot ring close to the inner boundary. This likely indicates that, in the absence of horizontal smoothing, our usual steady state (multiple hot rings at moderate heights) is not stable. See \S \ref{sec:stall} for more discussion.

We also test how the steady state depends on our assumed relaxation time (eq. \ref{eq:tauthermal}). We lengthen the relaxation time by a factor of $10$ overall and integrate for about 6 times longer. There is no appreciable difference in the steady state. This suggests that the thermal time only sets the overall timescale for dynamics, but not the final outcome. We also experiment with a  thermal time that increases inward (by multiplying the fiducial one by a factor of $(r/20 {\rm au})^{-1/2}$). We observe no difference in the final steady state. 
\begin{figure}
    \centering
    \includegraphics[width=0.48\textwidth]{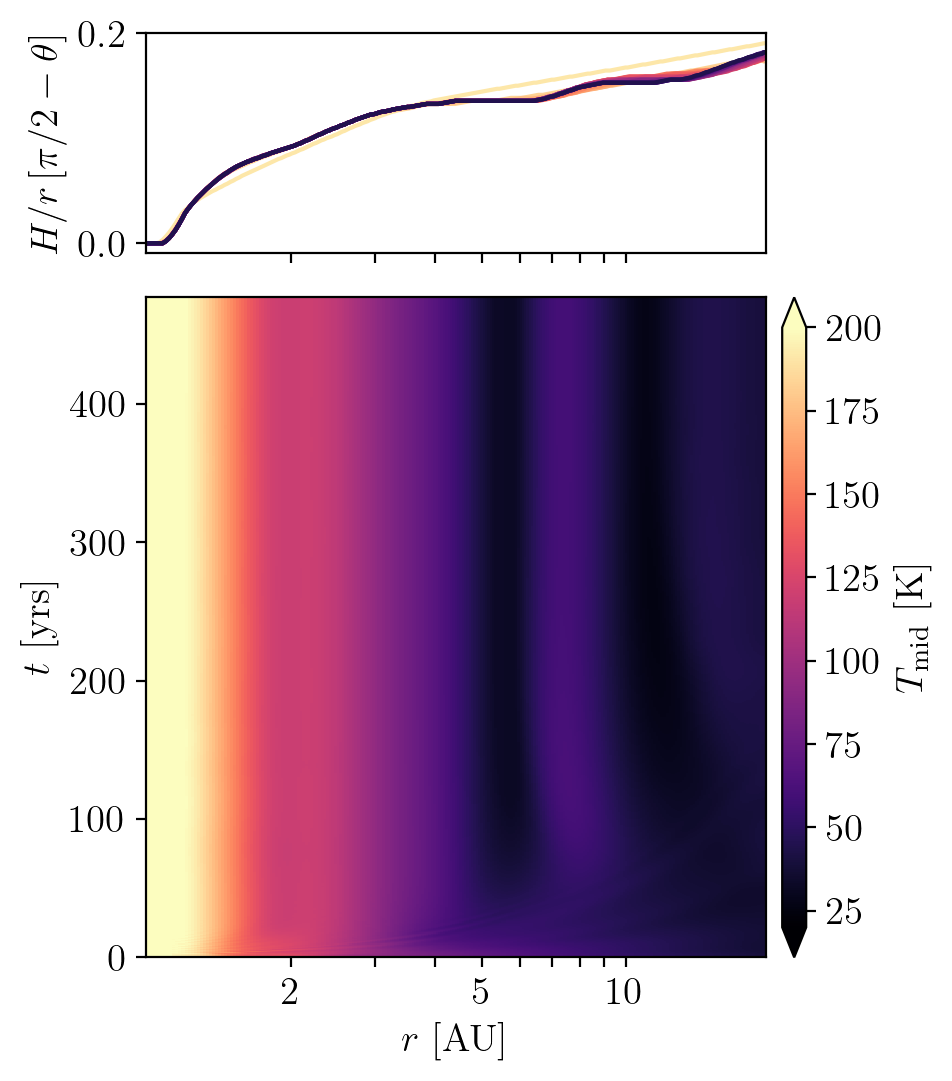}
    \caption{Similar to Fig. \ref{fig:fiducial} but for a disk that includes a "live" inner rim (eq. \ref{eq:rim}).  The inner rim is heated and puffed up under direct exposure of starlight. This casts a shadow to $\sim 7$ AU. Beyond this, the disk again develops hot rings and dark gaps.
    }
\label{fig:rimwave}
\end{figure}

Lastly, we experiment with a live inner rim (Fig. \ref{fig:rimwave}). Instead of  erecting a static  screen inward of 1AU, we simulate the response of a full disk that includes an inner cut-off. The disk surface-density profile is
\begin{equation}
    \Sigma(r) = \Sigma_0\Big( \frac{r}{r_0} \Big) ^{-1}\, \times \Big[1-e^{-\big(\frac{r}{ r_{\rm rim}}\big)^p}\Big]\,,
    \label{eq:rim}
\end{equation}
where we adopt $r_{\rm rim} = 3$AU and $p=8$ to emulate simulations in \citetalias{Pav}.\footnote{Special care must be taken near the inner edge where the thermal time is short and the integration time-step needs to be small. We artificially lengthen the thermal time there by a factor of $5/(\kappa_V \Sigma)$ where-ever $\kappa_V \Sigma \leq 5$.} As the inner rim is now exposed to direct starlight, it is heated to a temperature much above the \citetalias{CG97} values. This casts a long shadow out to $\sim 7$AU. But beyond the shadow, the disk is again able to form structures. The overall dynamics is similar to the fiducial case. 

In summary, the staircase steady state remains robust to changes in model parameters.

\section{Why do the thermal waves stall?}
\label{sec:stall}

One hallmark feature of the irradiation instability is that the perturbation propagates towards the star. This is found in the linear analysis of \citetalias{WL}, and confirmed by numerical integrations by \citet{Watanabe2008,Ueda2021,Okuzumi2022}. It is also seen in the early stages of our simulations (see Fig. \ref{fig:evol}). However, after a while, the waves in our simulations stall to form a staircase steady state. This is surprising. 

This stalling could be an artefact of our numerical implementations. Alternatively, it could be genuine, either as a result of the hydrodynamics (above cited studies all assume hydrostatic equilibrium), or that the staircase steady state is genuinely immune to the irradiation instability. Here, we discuss each in turn.

We adopt a number of short-cuts in the simulations to deal with the complicated thermal physics. They are not the cause for the stalling: the vertical isothermal assumption  is the same as that adopted by the study of \citetalias{WL} which finds travelling waves; the amount of horizontal smoothing impacts the positions of the hot rings but not the overall results (Fig. \ref{fig:comparison}); adopting either a static inner screen or a live inner rim leads to similar results; the  adoption of a temperature floor is motivated by physics (the 'back-lighting' effect), is confirmed by RADMC, and waves stall   even when we remove this floor. Lastly, our simulations do produce travelling thermal waves in the initial stage, and waves stall only after a period of growth. This suggests some nonlinear effects are at play.

We consider one nonlinear effect: fluid advection. According to the perturbation analysis of \citetalias{WL} (see their eq. B25), the thermal perturbation should propagate inward with a phase speed of  $v_{\rm phase} \sim - r/\tau_{\rm th}$, which is of order $0.1 v_{\rm kep} - 10 v_{\rm kep}$ for our simulations. A fluid flow with a similar velocity can in principle suppress the inward propagation by advecting the thermal energy away from the hot front. However, typical meridional speed we see at our steady state (Fig. \ref{fig:streamlines}) is only of order $10^{-3} c_s$ and falls short for the task.

Another nonlinear effect is changes to gas pressure at equilibrium. In the initial power-law state, gas pressure decreases outward monotonically. But as the thermal waves grow in amplitude, some gas is expelled from the hot ring and piles up ahead and behind it (Fig. \ref{fig:pgrads}).
This, coupled with the evolving temperature,  flattens the radial pressure profile ahead of a hot ring. In particular, we observe that wave stalling appears to occur when the pressure in this region develops a inflection point (but not a pressure maximum). However, we do not understand the causal connection between these two events and leave this for a future study.

Lastly, if the staircase state (the nonlinear state) is stable to further irradiation instability, it may explain why the wave stalls. We borrow directly results from \citetalias{WL}. They considered the irradiation instability of a power-law disk with a flaring index
\begin{equation}
\gamma \equiv {{d\ln (h/r)}\over{d\ln r}}\,, 
    \label{eq:gamma}
\end{equation}
where $h$ is the local scale height and $h/r = c_s/v_{\rm kep}$. This index equals $2/7$ for the conventional \citetalias{CG97} disk (eq. \ref{eq:CG97}), but we measure a much steeper flaring, $\gamma \sim 2$, at the front edges of our hot rings. The irradiation instability prefers waves with short wavelengths. This is because, when a disk is locally heated up, a shorter wavelength mode can produce a steeper optical surface (larger $dH/dr$), allowing the disk to intercept more starlight and  overcome the higher black-body loss. Eq. (31) of \citetalias{WL} states that, for a thermal perturbation of the form $X \propto e^{st + i k \ln r}$, the instability sets in when the wave-number 
\begin{equation}
k \geq k_{\rm min} = \sqrt{\frac{7}{\chi^2 - 8}}\, \chi^2 \gamma \sim 30 \left(\gamma\over 2\right)\, ,
\label{eq:growth}
\end{equation}
where $\chi = H/h$ and we have evaluated using $\chi = 4$. So for our case at hand ($\gamma \approx 2$), unstable waves have very short radial wave-lengths, $\sim r/k \sim 0.8 h \times \left(\frac{h/r}{0.04}\right)$. This is small compared to our typical smoothing length of $H \sim 4 h$. So at face value, the staircase state seems to be less, or even not, susceptible to the irradiation instability.

Some support for this conjecture is provided by a simulation with zero smoothing (Fig. \ref{fig:comparison}). In this case, it appears that the configuration of multiple hot rings is not stable and the disk continues to evolve until it builds up one very hot ring close to the inner boundary (at which point the boundary condition may come into play).   

To summarize, we provide arguments that the wave stalling is likely physical and is a result of nonlinear evolution. However, we are not certain of the actual mechanism for stalling. More detailed analysis is warranted. 


\section{Comments on Previous Works}
\label{sec:critique}

Our result, that smooth disks spontaneously develop into a staircase steady state, runs in direct contradiction to two recent claims by \citetalias{MF} and \citetalias{Pav}. Both studies have  simulated irradiated disks but found that their simulations produce smooth disks at steady state.

Here, we will examine their respective steady states in detail. In doing so, we bypass the issue of time-dependent radiative transfer, which are treated differently in our and their works. Moreover, for systems in steady state, the thermal equilibrium should be well characterized by Monte-Carlo codes like RADMC3d. So we have the `ground-truth' to bench-mark results of radiation hydro codes.

\subsection{Comments on \citet{MF}}\label{sec:MFK}

\begin{figure}
    \centering
\includegraphics[width=0.46\textwidth]{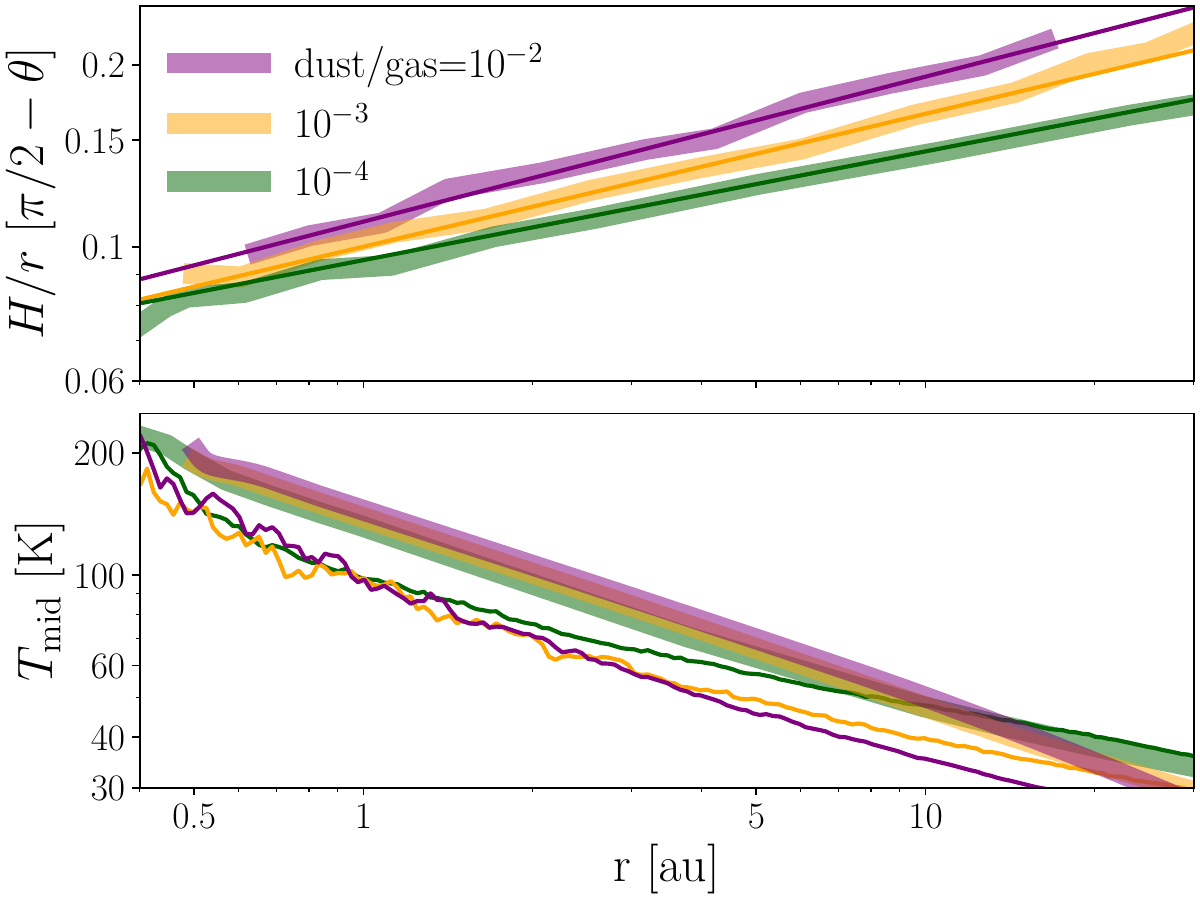}
    \caption{Comparison of the steady state reached in \citetalias{MF}  (thick lines, digitized from their Fig. 12 and 15) against those obtained from RADMC3d (thin lines). The different colors stand for the different dust to gas ratios, as considered in \citetalias{MF}. The RADMC3d input disks are adjusted to share similar density structures and identical optical surfaces (top panel) as those from \citetalias{MF}. However, the bottom panel shows that the midplane temperatures in the \citetalias{MF} disks are up to $40\%$ hotter than what RADMC3d predicts. This translates to a heating flux that is up to a factor of $4$ too large. Discrepancy between the two codes is worse  in the inner regions where the disks are optically thick to their own thermal radiation.     }
\label{fig:compareMK}
\end{figure}

In \citetalias{MF}, radiative transfer is treated by following only the first two moments of specific intensity (mean intensity and flux), with the higher moments truncated by way of the so-called Eddington tensor under M1 closure \citep{Levermore}. Such a technique allows them to capture radiative transfer in  the optically thick limit and, to a less accuracy, in the optically thin domain. They report that an initial staircase disk relaxes to a smooth, power-law steady state.

To examine the steady state that \citetalias{MF} obtained, we need their 2D temperature field, which Dr. Melon Fuksman has kindly provided. We also need the 2D density field, which we construct by digitizing the optical surface in their Figure 12 and assuming that the disk is vertically isothermal and is in hydrostatic equilibrium. The stellar parameters and the opacity law\footnote{\citetalias{MF} assumed that the opacity is contributed entirely by small grains. So it falls off steeply with photon wavelengths as $\kappa_\nu \propto \lambda^{-2}$. This will become relevant below.} are as provided by \citetalias{MF}, as is the surface density profile (which includes an inert disk inward of $0.4$au). 
We pay special attention to ensure that our disk shares the same optical surface as theirs,\footnote{We confirm that the result is only sensitive to the value of $H$ and the surface density (via the vertical optical depth), but not so much on the vertical density distribution. } sometimes by cropping all gas that lies above $H$.\footnote{This involves only a small amount of gas. So the surface density is largely unchanged.} With such a procedure, we can be assured that our disk shares the same amount of stellar heating as theirs. Lastly, we use RADMC3d to obtain the equilibrium temperature field for such a density field. A large number of photon packages ($5\times 10^8$) are used such that the temperature results converge even in very optically thick regions. 

Fig. \ref{fig:compareMK} shows the comparison for the three cases investigated by \citetalias{MF}. In every case, the  RADMC3d midplane temperatures fall significantly below those from \citetalias{MF}, by up to $40\%$. The discrepancy is  worse in the more optically thick inner regions and in disks that are dustier (and therefore more opaque). Satisfactory agreement exists only in one case: the outer part of the
disk that has the lowest opacity. In this part, the disk is optically thin to its own radiation. Interestingly, this is also the limit for which the two-moments method has been well calibrated \citep{Fuksman2021}.

The overall discrepancy in midplane temperature is significant. It means that, for whatever reason, the \citetalias{MF} disks receive up to 4 times too much heating flux, compared to what  RADMC determines.

To convince ourselves of the RADMC3d results, we analyze in detail (Appendix \ref{sec:opacity}) the disk thermal structure for different opacity laws and optical depths. 
In a nutshell, for a disk that is optically thick and has an opacity law as steep as that adopted by \citetalias{MF}, the disk's IR photosphere  lies well above the mm photosphere and deflects half of the illuminating flux back to space. As a result, the midplane becomes cooler than that with a grey-opacity.  RADMC3d correctly captures this process and we have confidence in its results.

We do not understand why the two-moments code transport too much flux to the midplane. But we can speculate on why it may matter for our problem at hand. In a disk with a cooler midplane, for instance, the disk immediately beyond the inner screen should become thinner by up to $20\%$. It will then be blocked from the star and suffers cooling collapse. At the same time, the  disk further out can now receive more sunshine. These changes may then lead to the formation of a stair-case. In other words, the hotter disks in \citetalias{MF}, which are in contradiction with RADMC, may have precluded the irradiation instability. 

In summary, while \citetalias{MF} claimed that the irradiation 
instability is suppressed by the combined effects of vertical thermal diffusion and fluid advection, our analysis casts doubt on their final equilibrium state, and by association, the validity of their claim. Calibrating the equilibrium state, especially for optically thick disks, may be a necessary (though not sufficient) safety guard for any radiation hydro codes.
 
\subsection{Comments on \citet{Pav}}
\label{subsec:pav}

\begin{figure}
    \centering
    \includegraphics[width=0.48\textwidth]{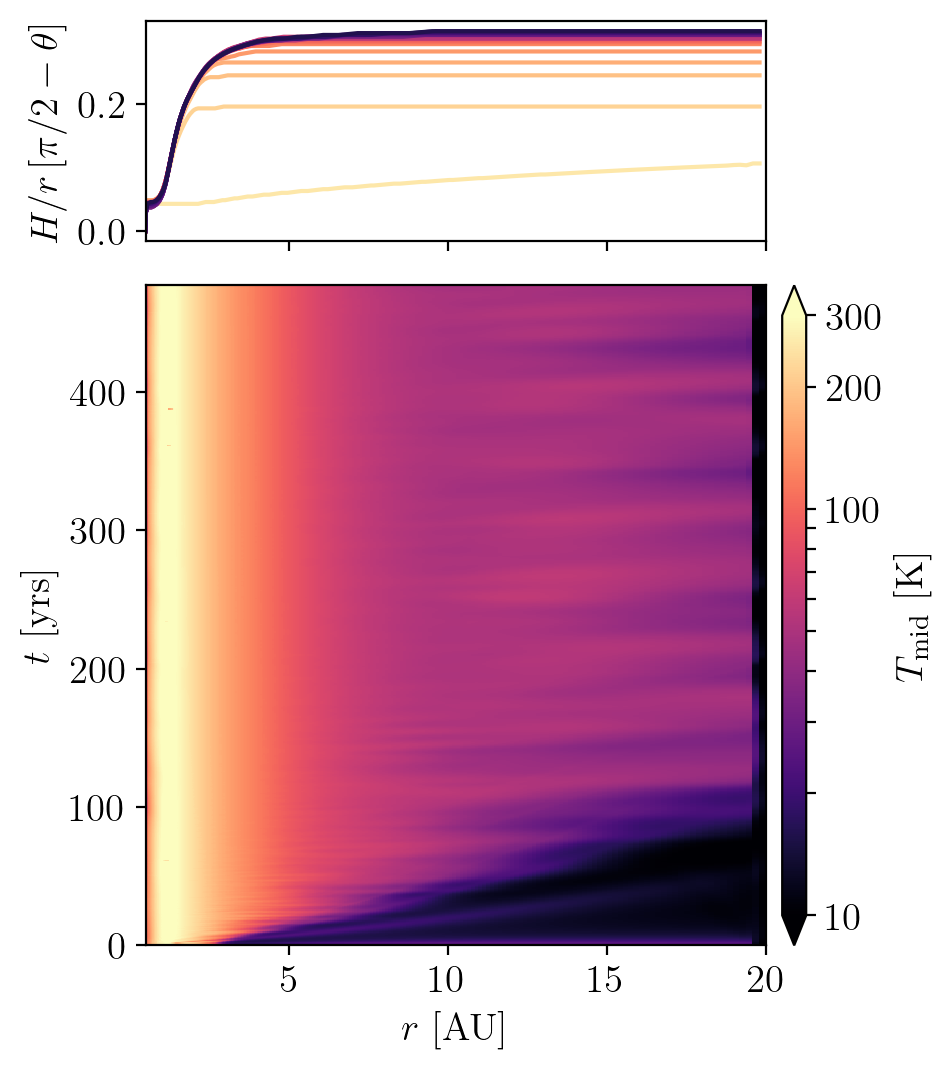}
    \caption{Similar to Fig. \ref{fig:rimwave} but for a case that is designed to reproduce the simulation of  \citetalias{Pav}.  The disk has a uniform initial temperature of $10\K$. The higher stellar luminosity here heats the inner rim up to such a high temperature that it casts a long shadow out to beyond 20au. This explains why \citetalias{Pav} did not find sub-structures.   
    }    \label{fig:shadow}
\end{figure}

\begin{figure*}[t]
    \centering
\includegraphics[width=0.95\textwidth]{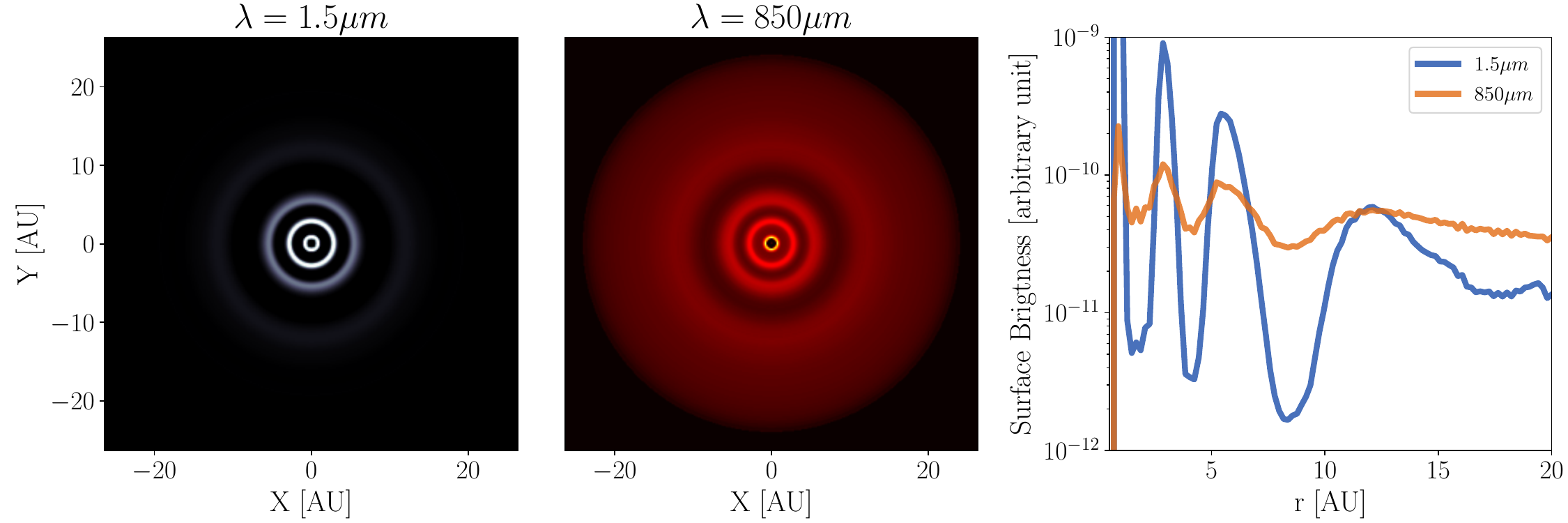} 
    \caption{A visual impression of the staircase steady state from Figure \ref{fig:fiducial}, in both $1.5\mu m$ scattered light image (left, linear scale), and $850\mu m$ thermal emission (middle, linear scale). The right panel shows the   surface brightness profiles in logarithmic scale.   Even though the surface density in this disk is smooth, prominent sub-structures are present in both wavelengths. Bright rings in scattered light correspond to where the disk sees the star (the "stair-risers"), and those in sub-mm correspond to where the disk is hot.}
\label{fig:pretty}
\end{figure*}

Using a different radiation hydro code (following only the first moment of specific intensity, and truncating higher moments by the diffusion approximation), \citetalias{Pav} also found that their disks reach a smooth temperature profile. 

Their study includes a live inner rim that is exposed to direct starlight. They also adopt a very high opacity, as well as a high stellar luminosity ($5L_\odot$, as opposed to $1 L_\odot$ in this study).

To examine how these may impact the outcome, we modify our simulation to emulate their runs:  $T_{*} = 6000 \, {\rm K}$, $R_* = 2 R_\odot$, such that $L_* = 5 L_\odot $; a uniform initial disk temperature of $10\K$; this is also now the floor temperature;  a radial domain that runs from $0.5$ to  $20$ au;  a gas surface density profile as in eq. \refnew{eq:rim} but with $\Sigma_0 = 200 \g/\cm^2$; a Planck opacity of $\kappa_V = 50 \cm^2/$g-gas  \citep[see][]{pav2020}. We do not include dust scattering, but it is unclear if it is included in their work. 

Our results are presented in Fig. \ref{fig:shadow}. Compared to our run shown in Fig. \ref{fig:rimwave}, the higher stellar luminosity (and to a lesser degree, the higher opacity) in this case warms up the inner rim to a higher temperature. It expands and can now intercept stellar fluxes up to a height of $z/r \sim 0.3$. This shadows the entire disk to beyond the  integration boundary ($20$AU). Our plot resembles the results in  Fig. 8 of \citetalias{Pav}. Similarly to what we find here, their outer disk appears very cold and lies largely in shadow. This explains why they do not find substructures -- much of the disk is not irradiated. 

It is possible that, were they to adopt a lower stellar illumination, or a lower dust opacity, or a different shape of the inner rim, the dynamics and the steady state could change. 

\begin{figure}
    \centering
    \includegraphics[width=0.45\textwidth]{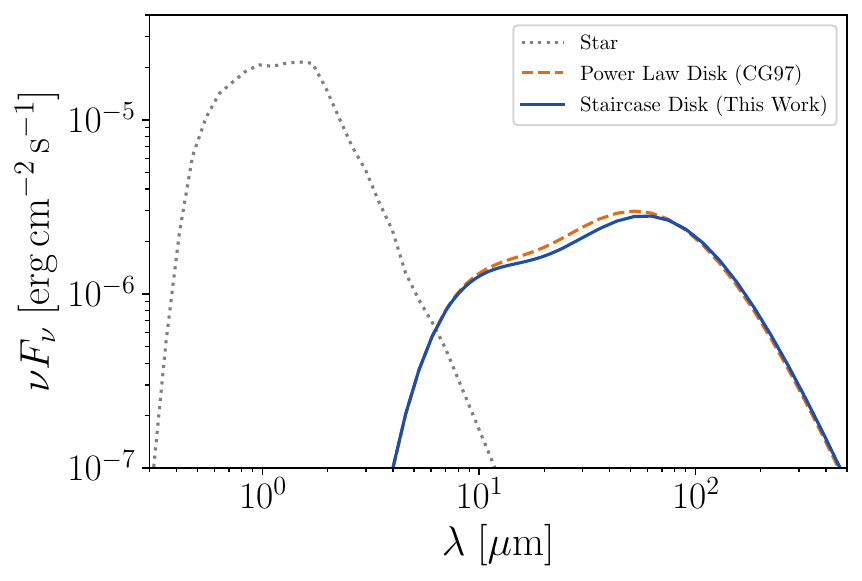} 
    \caption{Spectral energy distributions for our staircase disk (from Fig. \ref{fig:fiducial}, solid curve) and the same disk but with a power-law temperature profile (eq. \ref{eq:CG97}, dashed curve), both measured at 150pc. These two disks intercept and reprocess similar amounts of stellar heating, so their SEDs are nearly indistinguishable.
    The dotted curve is the stellar SED. }
\label{fig:sed}
\end{figure}

\section{Observable Consequences}
\label{sec:observable}

While our conclusions are far from  being definitive, we muse about a few observational consequences of the staircase disks. 

We first ask whether one can recognize a staircase disk if only knowing the spectral energy distribution (SED). We post-process, both the initial (power-law like, flared) and the staircase final state, through RADMC3d (see details in \S \ref{subsec:radmc}) to produce their respective SEDs. These are shown in Fig. \ref{fig:sed}. There is little difference. This is as expected because both disks intercept and reprocess similar amounts of stellar flux.

In resolved images, on the other hand, Fig. \ref{fig:pretty} shows that substructures like bright rings and dark gaps are clearly visible. These are reminiscent of those discovered abundantly by high resolution  observations, both using adaptive optics in the optical/IR \citep[e.g.] []{Garufi2016,Avenhaus2018}, and using interferometry in sub-mm by ALMA \citep[e.g.][]{ALMA2015,Huang2018}. 
In scattered light,\footnote{Here, we turn on dust scattering in RADMC3d during post-processing.}, the hot rings can shine up to 100 times brighter than the dark zones. The millimeter map shows a more muted contrast between the substructures, though dust drift may amplify it further.
Here, we are focused on the region inward of $20 {\rm AU}$, where there has been little observational evidence regarding the presence of sub-structures. This may be related to instrument resolution, and it is of interest to note that recent works have discovered substructures also in compact disks \citep[smaller than $50 {\rm AU}$ in size,][]{Long2018,Zhang2023}. 

\begin{figure}
    \centering
    \includegraphics[width=0.45\textwidth]{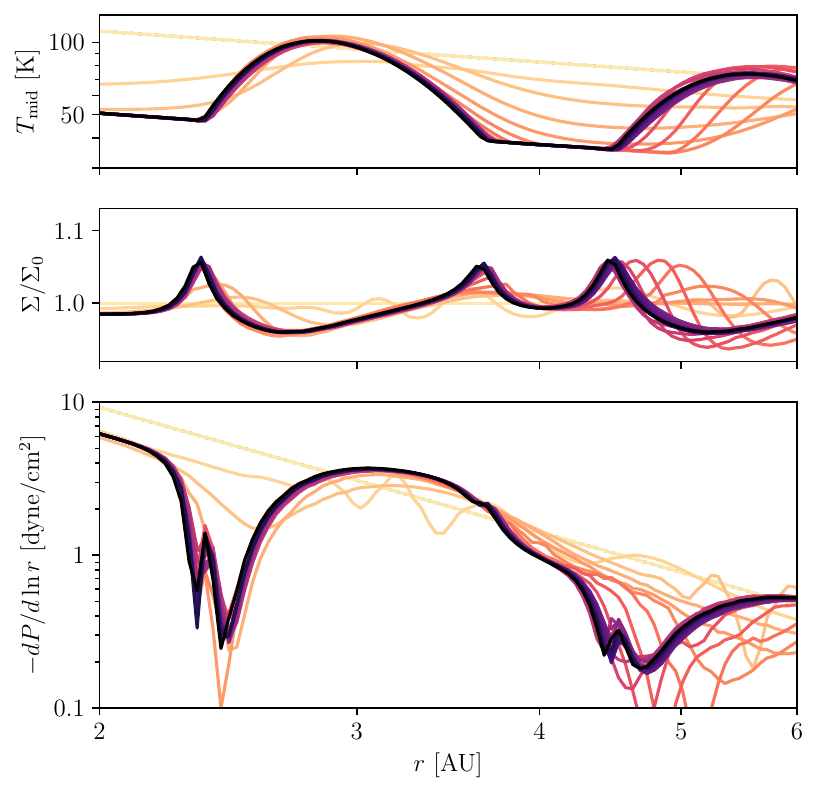}
    \caption{The radial pressure gradient (lower panel) is strongly modified by the presence of substructures. The solid lines show the 
    evolution of our fiducial case in the region around the first hot ring, with the line colors same as in Fig. \ref{fig:evol}.  
    The top panel also shows the midplane temperature, and the middle the ratio between surface density and its initial value. As the disk approaches equilibrium, the pressure gradient approaches zero near the front side of the hot ring. Grains will tend to concentrate here. The disk may also become unstable to Rossby wave instability or baroclinic instability.
    }
    \label{fig:pgrads}
\end{figure}

Lastly, we are curious to know if the radial pressure gradient in these stair-case disks can impact  dust drift. Fig. \ref{fig:pgrads} shows that, the pressure gradient near the front sides of the hot rings is weakened by almost two orders of magnitude, from values for smooth disks. This drastically reduces (though not completely stalls) the radial drift of all dust grains.   Dust may become concentrated near these regions. This evolution is interesting to explore.

There are also other unexplored consequences. The unusual pressure gradient near the hot rings may provide breathing grounds for the baroclinic instability \citep[see, e.g.][]{klahr2004}, or the Rossby wave instability \citep{lovelace1999}.
Vortices and/or spirals may form. While these instabilites do not manifest in our 2D simulations, they should be examined in 3D.

\section{Summary}
\label{sec:conclusion}

A passively irradiated disk abhores a smooth profile. We have
known, since the pioneering study of \citet{Dullemond2004}, that a passive disk can change shape and alter the stellar irradiation it receives. As is well documented for Herbig Ae/Be disks, the inner disk rim can puff up and block the region beyond it. In this work, we show that every part of a passive disk can partake in the choreography. 

As a natural consequence of the irradiation instability, an initially featureless disk spontaneously morphs into a `staircase' form, with hot rings that intercept the lion's share of the stellar irradiation, and dark zones hidden in their shadows. Each of the hot rings is analogous to a puffed-up inner rim. These sub-structures form after a few thermal time \citepalias[$\sim 10^5$ yrs in the inner region,  see Fig. 2 of][]{WL}, with the inner ones congealing into a static form first. Their locations depend on a number of parameters, including the initial conditions.

The existence of such a steady state is surprising, given that previous works which assume  hydrostatic equilibrium have found only inward travelling waves. We examine a few possible causes for the stalling. We argue that the stalling is likely genuine and the result of nonlinear evolution. But we have yet to understand its physical cause.

Major caveats exist for our work. Although our staircase steady states appear to be  in thermal and dynamical equilibria, we fall short of showing how, in reality, such a state is  reached. To do so, one needs to conduct simulations with realistic radiative transfer.  Along the same vein, the disagreement between our work and those of \citet{MF,Pav} is unlikely to be fully resolved until better treatments of radiative transfer are introduced.  Another major caveat is that our 2D simulations precludes non-axisymmetric hydro and thermal instabilities. These latter may qualitatively alter the picture. 

But if the staircase phenomenon is real, it has many interesting implications. It would offer an explanation for the prevalence of gaps and rings in observed disks. It could alter the physical state of the disk and affect processes such as the condensation of volatile species,\footnote{The locations for icelines would be different from the traditional picture. As the disk temperature is not  monotonic with radius, there may even be multiple icelines for the same species.} the radial dust drift, and the dust vertical wafting. It may introduce other hydro instabilities that are absent in smooth disks, and possibly affect disk accretion. The processes  of planet formation and migration could all be affected. In summary, a passive disk can be quite an interesting place.

\bigskip

\section*{Acknowledgements} 
The authors thank Eugene Chiang, Yan-fei Jiang, Francios Foucart, Zhaohuan Zhu, Xuening Bai, Shoji Mori, Ariel Amaral and an anonymous referee for helpful conversations, and Dr. Melon Fuksman for providing data. TK and YW acknowledge funding from NSERC; TK acknowledges further funding from the Walter C. Sumner Memorial Fellowships, and the Dunlap Institute. YL acknowledges NASA grant 80NSSC23K1262.

\bibliography{refs}

\appendix

\section{Midplane temperature and Opacity}
\label{sec:opacity}

Here, we present a brief study on the disk midplane temperature for different opacity laws, when at thermal equilibrium. This is an aspect of the irradiated disks that is not dealt with in previous literature \citep[see,e.g.][]{CG97,dalessio1998}. Typically in these works, one only considers photons at two frequency ranges (stellar radiation and disk thermal radiation). This is inaccurate in cases where the opacityy is a steep function of photon frequency.

For definiteness, we adopt the following density profile 
\begin{equation}
\rho (r,z) = {{\Sigma(r)}\over{\sqrt{2 \pi} h(r)}} \exp{\left[- {{z^2}\over{2 h(r)^2}}\right]}\, ,
\label{eq:fakerho}
\end{equation}
and the same disk parameters as in \citetalias{MF}:   $h(r)/r = 0.0248 (r/{\rm AU})^{0.275}$, and $\Sigma(r)= 200 \g/\cm^2 (r/{\rm AU})^{-1}$. 

In the simple case of a grey opacity law, $\kappa = {\rm const}$ at all wavelengths (the case we consider in the main text),\footnote{More generally, the opacity only needs to be constant in any wavelength that the disk radiates.} the disk is vertically isothermal below the optical surface, and radiates as a blackbody from the photosphere, defined as where the vertical optical depth $\tau_z = 1$. We can balance the heat gained from the hot dust grains on the optical surface, with the local blackbody heat loss, to obtain the temperature for this layer to be (eq. 10 in \citetalias{WL}): 
\begin{equation}
T_{\rm mid}(r) = \left[{1\over 2} {d\over{d\ln r}} \left({H\over r}\right)\right]^{1/4}\, \left({{R_*}\over{r}}\right)^{1/2}
T_* \, .
\label{eq:simpleT}
\end{equation}
Note that this expression is independent of surface density as long as the disk is vertically optically thick. 

Now let us consider an opacity of the form 
\begin{equation}
    \kappa_\lambda = \kappa_0 \left({\lambda\over{10 \mu m}}\right)^{\alpha}\, , \label{eq:kappa}
\end{equation}
with $\alpha \leq 0$. For instance, the opacity laws adopted by \citetalias{MF} and \citet{Pav} are contributed entirely by small grains and so fall off steeply with wavelength as $\alpha \sim -2$ for $\lambda \geq 10\mu m$. For easy model comparison, we adopt a constant opacity at short wavelengths ($\lambda < 10\mu$m) as $\kappa_\lambda = 10 \cm^2/\g-gas$. Such a choice guarantees the same optical surface $H$ (and therefore heating) for all models, and emphasizes the role played by the opacity law. 

A steep opacity law presents an interesting issue. We first assume that the disk is optically thick in the vertical direction to its own radiation.

\begin{figure*}[h]
    \centering
\includegraphics[width=0.95\textwidth]{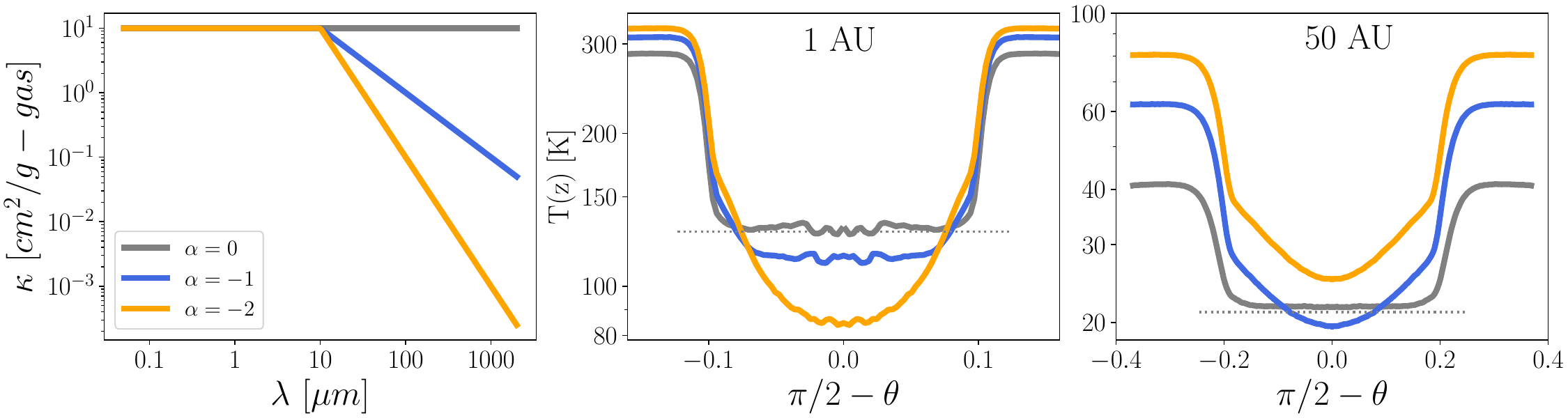}
    \caption{We confirm with RADMC that a steep opacity law as that adopted in \citetalias{MF} yields a colder $T_{\rm mid}$, compared to one where the opacity law is grey. The latter satisfies eq.  \ref{eq:simpleT} (dashed lines).  All disks have the same density profile (eq. \ref{eq:fakerho}), the same optical opacity and the same optical surface ($H$). The 1AU region is optically thick to all radiation; while 50AU is thin in some cases. }
\label{fig:tmid_kappa}
\end{figure*}

The surface that intercepts the reprocessed radiation from the optical surface (call it `IR' light) is at $\tau_{\rm IR} = 1$ (IR photosphere). It lies above the surface of disk thermal radiation (`mm' photosphere, $\tau_{\rm mm} = 1$). The former layer effectively deflect part of the illuminating flux to vacuum, allowing less flux to pass through to the column.  If the opacity law is sufficiently steep, there can be multiple such deflection layers. As a result, a more negative $\alpha$ means a cooler mid-plane. This behaviour is demonstrated in Fig. \ref{fig:tmid_kappa}, where we use RADMC3d to trace the temperature structure.

On the other hand, this effect is reversed for the optically thin regions, typical of the outer disk. Here, a more negative $\alpha$ brings about a hotter midplane, because the grains are less efficient radiating at long-wavelength. 

In conclusion, at the same stellar illumination, the midplane temperature of an optically thick disk depends on the column density and the opacity law. In the case where $\alpha$ is substantially negative (e.g., when small grains dominate the opacity), the midplane temperature cools with increasing column density. The conventional picture of a vertically isothermal disk is strictly true only for a grey opacity-law. This physics is correctly captured by the RADMC3d code.

\section{Upper Boundary Condition And Waves} \label{app:bc}

Gas density in an isothermal disk declines sharply away from the midplane. This creates numerical issues, especially for simulations that extend to more than a couple scale heights. Typically a density floor is introduced to avoid dealing with vacuum. But it also brings with it other undesirable consequences. We find that the upper boundary (where density floor is encountered) continuously excite waves. We have adopted special boundary conditions (see main text) to partially alleviate the problem. 
Here, we gauge how well this performs by comparing runs with and without irradition. The boundary effects are present in both cases, so the difference between the two is purely due to irradiation.

\begin{figure}[h]
    \centering
    \includegraphics[width=0.6\textwidth]{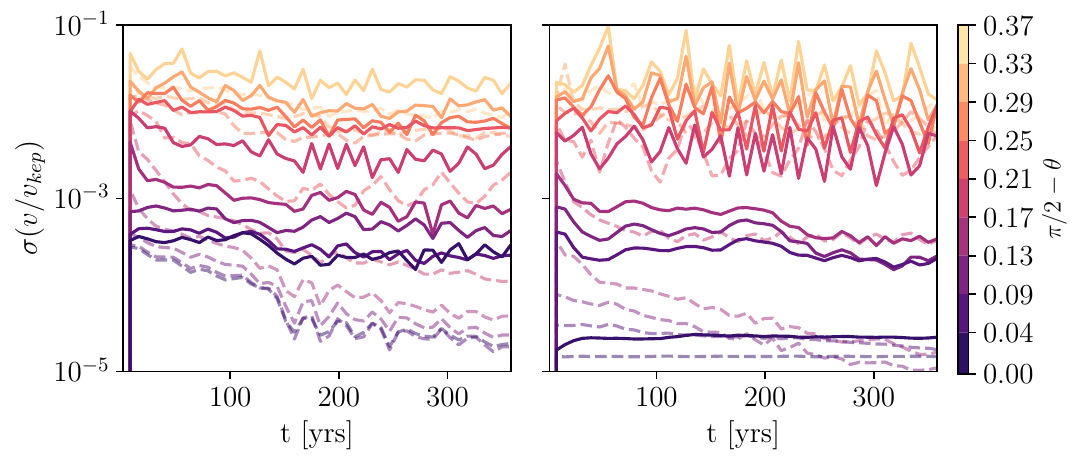}
    \caption{Standard deviation of  $(v_r/v_{\rm kep})$ (left) and $(v_\theta/v_{\rm kep})$ (right) across the radial domain, calculated at a given altitude ($z/r$, color bar to the right). The purely hydro case is shown in dashed lines, and the fiducial case in solid lines. The upper disks maintain large velocity dispersions in both simulations, while the lower disks  (below the optical surfaces) calm down over time to much smaller values. Some meridional motion remains in the irradiated case.
    }\label{fig:veldisp}
\end{figure}

Fig. \ref{fig:veldisp} displays the velocity dispersions at various heights above the midplane. All latitudes above the optical surface ($z/r \geq 0.17$) exhibit fast fluid flow, as a result of the imperfect boundary condition (also see Fig. \ref{fig:streamlines}). Latitudes nearer the midplane calm down with time, with the purely fluid disk calming faster (also see Fig. \ref{fig:bc}). We conclude that our imperfect upper boundary does not affect the main region (below the optical surface), but meridional velocites above the optical surface should not be trusted.

\end{document}